\documentclass[aps,pra,twocolumn,showpacs,superscriptaddress]{revtex4}
\usepackage{graphicx}
\usepackage{amsmath}
\usepackage{amssymb}
\usepackage{color}
\newcommand{\ket}[1]{\ensuremath{\left|{#1}\right\rangle}}
\newcommand{\bra}[1]{\ensuremath{\left\langle{#1}\right |}}

\newcommand{\oper}[1]{\boldsymbol{\mathsf{#1}}}

\newcommand{\rrho}{\ensuremath{\boldsymbol{\rho}}}
\newcommand{\W}{\mathcal{E}}
\newcommand{\qq}{\boldsymbol{q}}
\newcommand{\bq}{\bar{\boldsymbol{q}}}

\newcommand{\Int}{\int\!\!}
\graphicspath{{converted_graphics/}}

\begin{document}
\title{Klyshko's Advanced-Wave Picture in Stimulated Parametric Down-Conversion with a Spatially Structured Pump Beam}

\author{M. F. Z. Arruda}
\affiliation{Departamento de F\'isica, Universidade Federal de Santa Catarina, Florian\'opolis, SC, 88040-900, Brazil}
\affiliation{Instituto Federal do Mato Grosso, Sorriso, MT, 78890-000, Brazil}

\author{W. C. Soares}
\affiliation{Departamento de F\'isica, Universidade Federal de Santa Catarina, Florian\'opolis, SC, 88040-900, Brazil}
\affiliation{N\'ucleo de Ci\^encias Exatas - NCEX, Universidade Federal de Alagoas, Arapiraca, AL, 57309-005, Brazil}

\author{S. P. Walborn} 
\affiliation{Instituto de F\'isica, Universidade Federal do Rio de Janeiro, Caixa Postal 68528, Rio de Janeiro, RJ, 21945-970, Brazil}

\author{D. S. Tasca} 
\affiliation{Instituto de F\'isica, Universidade Federal Fluminense, Niter\'oi, RJ, 24210-346, Brazil}

\author{A. Kanaan}
\affiliation{Departamento de F\'isica, Universidade Federal de Santa Catarina, Florian\'opolis, SC, 88040-900, Brazil}

\author{R. Medeiros de Ara\'ujo}
\affiliation{Departamento de F\'isica, Universidade Federal de Santa Catarina, Florian\'opolis, SC, 88040-900, Brazil}

\author{P. H. Souto Ribeiro} 
\affiliation{Departamento de F\'isica, Universidade Federal de Santa Catarina, Florian\'opolis, SC, 88040-900, Brazil}

\date{\today}
\begin{abstract}
The advanced-wave picture is ``... an intuitive treatment of two-photon correlation with the help of the concept of an effective field acting upon one of the two detectors and formed by parametric conversion of the advanced wave emitted by the second detector ..." [A. V. Belinskii and D. N. Klyshko, JETP 78, 259 (1994)]. This quote from Belinskii and Klyshko nicely describes the concept of the advanced-wave picture; an intuitive tool for designing and predicting results from coincidence-based two-photon experiments. Up to now, the advanced-wave picture has been considered primarily for the case of an ideal plane-wave pump beam and only for design purposes. Here we study the advanced wave picture for a structured pump beam and in the context of stimulated emission provoked by an auxiliary input laser beam. This suggests stimulated parametric down-conversion as a useful experimental tool for testing the experimental sets designed with the advanced-wave picture. We present experimental results demonstrating the strategy of designing the experiment with advanced-wave picture and testing with stimulated emission.
\end{abstract}
\maketitle
\section{Introduction}

Optical correlations from spontaneous parametric down-conversion (SPDC) have been largely used to experimentally investigate fundamental aspects of quantum mechanics and to implement quantum information protocols. These correlations can be explored using several optical degrees of freedom, and are measured at the single-photon level by detecting the down-converted photon pairs with mode analyzers and coincidence electronics.

In a two-photon experiment, the coincidence-count distribution $C(\phi_1,\phi_2)$ is obtained by projecting photons $1$ and $2$ onto the optical modes $\phi_1$ and $\phi_2$. The coincidence-count distribution is proportional to the joint probability $\mathcal{P}(\phi_1,\phi_2)$ for the detection of the photon pair in these optical modes. Among the optical degrees of freedom for which down-converted photons may exhibit correlations, transverse modes are an interesting subject of study \cite{walborn10a}, as in this case the two-photon correlations extend over a wide range of spatial modes $\{\phi_1,\phi_2 \}$.

Even though the structure of the two-photon spatial correlations $\mathcal{P}(\phi_1,\phi_2)$ may be intricate \cite{Miatto12a,Miatto12b,Walborn12}, D. N. Klyshko developed a simple method, introduced in 1988, for describing these correlations when considering a SPDC source excited by a plane-wave pump-photon \cite{Klyshko88}. Instead of considering the {\it non-local} joint detections that take place in a real-life two-photon experiment, Klyshko's advanced-wave picture (AWP) is based on a {\it prepare-and-measure scenario}. In this scenario, the {\it detection event} at one of the detectors is replaced with (or thought of as) an {\it emission event}. Then, in place of detecting a photon (say photon $1$) in the spatial mode $\phi_1$, in the AWP one is effectively preparing a photon with transverse spatial mode $\phi_1$. This prepare-and-measure scheme is assigned with a conditional probability $\mathcal{P}_{AWP}(\phi_2|\phi_1)$ of detecting the emitted photon in the spatial mode $\phi_2$ after some propagation,  given that it was prepared in spatial mode $\phi_1$.  Klyshko's AWP is constructed in such a way that this single-photon prepare-and-measure probability equals the two-photon joint probability for the detection of photon $2$ in spatial mode $\phi_2$ given that its correlated down-converted partner photon was detected in spatial mode $\phi_1$: $\mathcal{P}_{AWP}(\phi_2|\phi_1)=\mathcal{P}(\phi_1,\phi_2)$. Being based on the preparation and measurement of an {\it advanced wave}, the AWP can be cast in terms of a classical optics experiment in which the detected {\it intensities} of an optical field are proportional to the SPDC joint probabilities measured at the single-photon level.

In 1994, Belinskii and Klyshko  \cite{Belinskii94} theoretically analyzed two-photon image and diffraction effects, where the conditions for the formation of a two-photon image and diffraction were predicted using a classical optics setup based on the AWP. This was an interesting demonstration of the usefulness of the AWP for designing experiments and predicting results.
Two-photon imaging and diffraction are optical effects observed in the spatial distribution of the joint detection of the down-converted optical fields, and rely on spatial-mode correlations between them. These two-photon effects were experimentally demonstrated a few years later in a variety of experiments \cite{SoutoRibeiro94,pittman95,strekalov95,pittman96a,SoutoRibeiro96,monken98a}. 
\begin{figure}
\centering
\includegraphics[width=\columnwidth]{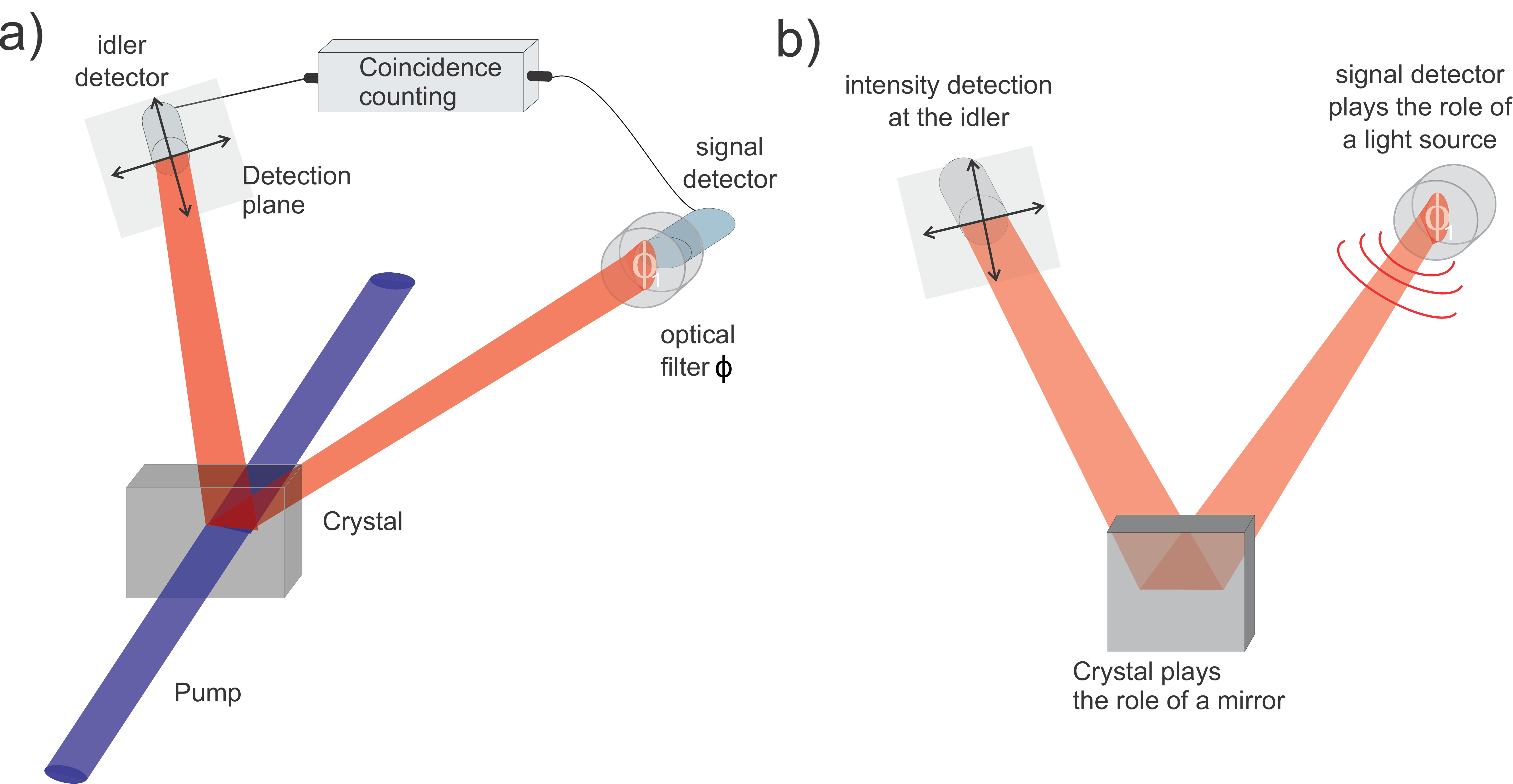}
\caption{a) Usual SPDC scenario and b) the advanced-wave picture (AWP).}
\label{fig:spdc_awp} 
\end{figure} 

A feature of the AWP is that this advanced wave emitted from detector-1's location is analogous to the temporal reversion of the down-converted field 1, implying that it is emitted towards the SPDC source. Then, ``...the advanced wave is effectively reflected by the wavefronts of the pump wave inside the crystal, {\it i.e.}, the thin crystal serves as a mirror for the advanced wave..." (Belinskii and Klyshko \cite{Belinskii94}). Finally, the reflected advanced wave is propagated from the SPDC source to detector $2$, taking into account all linear optical elements in place.

For a plane-wave pump considered by Klyshko \cite{Klyshko88,Belinskii94}, the non-linear crystal is simply replaced with a plane mirror of infinite transverse size. In this case, the law of reflection on a specular surface accounts for the perfect transverse wave-vector correlations between photon pairs created from a pump photon with well defined wave-vector. Pump beam phase curvatures lead to a curved, transversely infinite mirror, and have also been considered in the context of AWP \cite{pittman96a}.  More recently, an experimental demonstration of the equivalence between two-photon image and the AWP was carried out using a camera-based coincidence system \cite{aspden14}, and a classical optics ``prepare-and-measure"  experiment was used to predict novel orbital angular momentum correlations in SPDC \cite{Zhang14}.

In the majority of SPDC experiments, the pump beam is a well collimated zero-order Gaussian beam, often approximated in theory by a plane wave. Although this approximation reveals itself useful for understanding a huge variety of phenomena, a number of experiments have been performed that exploit the spatial structure of the pump beam.  These include the control of two-photon interference \cite{fonseca99b,fonseca00,kim00,walborn03a,walborn03b,nogueira04}, manipulation of correlations in orbital angular momentum \cite{mair01,walborn04a,Romero12}, creation of novel optical vortex structures \cite{gomes09a,gomes11}, violation of Bell's inequalities \cite{yarnall07a}, increase of spatial entanglement \cite{Walborn12,walborn07b} and exploration of higher-order quantum correlations \cite{gomes09b,Walborn11}. These experiments fall outside the usual treatment of the AWP. In this paper, we consider a more general Parametric Down-Conversion (PDC) experiment in which the pump beam has arbitrary spatial structure. We show that the two-photon coincidence distribution can be compared to the result of the propagation of the advanced wave through an optical element with profile equivalent to the pump profile. In this treatment, the crystal can be considered as an \textit{optically addressed spatial light modulator}, where both phase and amplitude are controlled by the pump laser beam. 

In order to illustrate the concepts above, we use Stimulated PDC (StimPDC), a cavity-free parametric amplifier that has been used to demonstrate image and coherence transfer \cite{Wang91,PRA-ImageCoh,Cardoso2018}, phase conjugation \cite{PRL-PhaseConj} and orbital angular momentum conservation \cite{PRA-OAM}. Stimulated emission has also been used for investigating the spectral properties of PDC \cite{eckstein2014,fang2014}. In StimPDC, the creation of photons on one of the down-converted fields is stimulated by an auxiliary laser. We show here that the AWP is also applicable to StimPDC, making it a helpful experimental tool for aligning and designing SPDC experiments that explore spatial correlations.

The paper is organized as follows. In section \ref{sec:parax}, we derive the equation describing the propagation of an optical field (transverse spatial amplitude) through a set of three linear optical devices in the paraxial approximation. It provides the classical model behind the AWP. In section \ref{sec:QS}, we derive the coincidence counting rate amplitude for SPDC, when signal and idler photons propagate through linear optical devices. We show that it is isomorphic to the classical field amplitude derived in section \ref{sec:parax}, provided that one of the photons is projected onto a mode having the same angular spectrum as the classical input field with a conjugated phase. In section \ref{sec:stim}, we derive the idler intensity distribution in StimPDC and make a connection with
the AWP classical model and the coincidence measurements, discussed in sections \ref{sec:parax} and \ref{sec:QS}, respectively. In section\ref{sec:experiments}, we present three StimPDC experiments correctly described by the theory presented in section \ref{sec:stim}. Finally, in section \ref{sec:design}, we provide insights on how the
StimPDC and the AWP can be useful to design coincidence counting experiments.

\section{Paraxial wave propagation}
\label{sec:parax}

The propagation of the electric field through a linear optical system can be described, in a very general way, by the following input/output relation \cite{saleh91}
\begin{equation}
\phi_{out}(\qq)=\int d\qq'\,H(\qq',\qq)\phi_{in}(\qq')\ ,
\label{eq:1}
\end{equation}
where $\phi(\qq)$ is the field's \textit{angular spectrum}, which is the Fourier transform of the field's amplitude profile $\mathcal{E}(\rrho)$:
\begin{equation}
\phi(\qq)=\frac{1}{2\pi}\int d\rrho\,\mathcal{E}(\rrho)e^{-i\qq\cdot\rrho}\ ,
\label{eq:angular-spec}
\end{equation}
$\qq=(k_x,k_y)$ are the transverse wave vector coordinates and $\rrho=(x,y)$ are its conjugate position coordinates. $H(\qq',\qq)$ is the impulse-response function of the optical device in the transverse wave vector domain, which we henceforth refer to simply as transfer function. All integrals have limits $-\infty$ and $\infty$ unless otherwise noted.

Let us consider a few particular cases that are going to be useful further in this article. The first one is the free paraxial propagation over a distance $z$:
\begin{equation}
H_z(\qq',\qq)=\delta(\qq'-\qq)\exp\left[-i\frac{q^2}{2k}z\right],
\label{eq:freeprop}
\end{equation}
where $k=\sqrt{q^2+k_z^2}$ is the wave number and $k_z$ is the $z$ component of the wave vector $\boldsymbol{k}$. The above transfer function, after integration in Eq. (\ref{eq:1}), simply multiplies the initial angular spectrum by a phase factor.

The second case is the amplitude-and-phase mask. It is described in the position space by a function $T(\rrho)$ relating the field $\mathcal{E}_+$ immediately after the mask to the field $\mathcal{E}_-$ immediately before:
\begin{equation}
\mathcal{E}_+(\rrho) = T(\rrho)\,\mathcal{E}_-(\rrho)\ .
\label{eq:mask}
\end{equation}
In transverse momentum space, the equation above is written in terms of the function $t$, the Fourier transform of $T$:
\begin{equation}
\phi_+(\qq)=\int d\qq'\,t(\qq'-\qq)\,\phi_-(\qq')\ ,
\end{equation}
from which $t(\qq'-\qq)$ is the mask's transfer function.

\begin{figure}
\centering
\includegraphics[width=0.9\columnwidth]{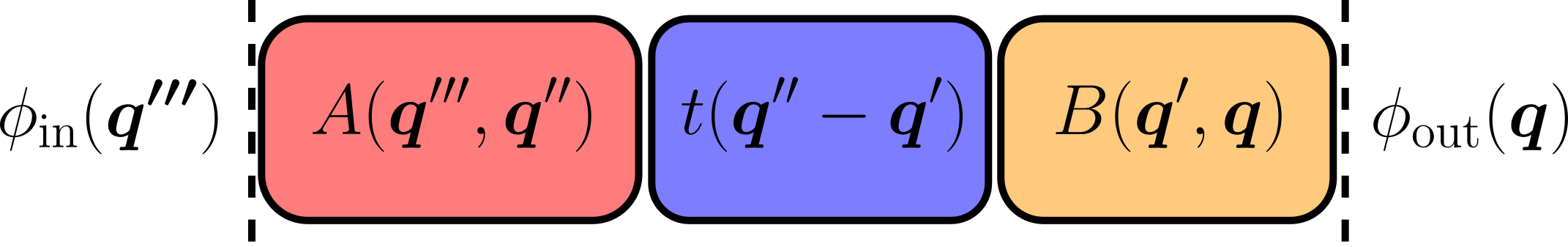}
\caption{Paraxial propagation through two arbitrary optical systems ($A$ and $B$) and an amplitude mask of transfer function $t$.}
\label{fig:awp_concept} 
\end{figure} 

Finally, we examine an optical system composed of three subsystems as shown in Fig. \ref{fig:awp_concept}, where, in between two subsystems of arbitrary transfer functions $A$ and $B$, we find a mask as described by Eq. (\ref{eq:mask}). The output \emph{angular spectrum} of this optical system is related to its input by
\begin{align}
\phi_{out}(\qq) = \int & d\qq' d\qq'' d\qq''' 
B(\qq',\qq)\,t(\qq''-\qq') \nonumber\\
& \times A(\qq''',\qq'')\phi_{in}(\qq''')\ .
\label{eq:in-out-q}
\end{align} 
The output \emph{amplitude} in spatial coordinates is therefore given by:
\begin{align}
\mathcal{E}_{out}(\rrho) = \frac{1}{2\pi}\int & d\qq d\qq' d\qq'' d\qq'''  e^{i\qq\cdot\rrho} B(\qq',\qq)t(\qq''-\qq')\nonumber \\
& \times  A(\qq''',\qq'')\phi_{in}(\qq''')\ ,
\label{eq:in-out-E}
\end{align}  
where $\phi_{in}$ can yet be written in terms of $\mathcal{E}_{in}$, allowing us to express the output field in terms of the integral transform of the input field. The last equation will be of particular help to build the Advance Wave Picture for both spontaneous and stimulated PDC scenarios. Note that $A$ and $B$ are completely arbitrary paraxial transfer functions that may comprise free propagation and/or any spatial modulations possibly introduced by the optical subsystems.

\section{Two-photon quantum state generated by SPDC}
\label{sec:QS}

SPDC is a wave-mixing process involving three fields: pump, signal and idler. The pump is usually a laser beam, while signal and idler are weak fields, due to the low conversion efficiency of the spontaneous process. In the monochromatic, paraxial and thin-crystal approximations, the two-photon state produced by SPDC is well described by \cite{monken98a,walborn03a,Costa-Moura10,walborn10a}
\begin{equation}
\ket{\psi}=\ket{\textrm{vac}}+C\int d\qq_1 d\qq_2\ 
v(\qq_1+\qq_2)\ket{1;\qq_1}\ket{1;\qq_2}, 
\label{eq:state}
\end{equation}
where $\ket{1;\qq}$ represents a single-photon state in the mode with transverse momentum $\qq$ and $v(\qq)$ is the normalized angular spectrum of the pump beam at the exit plane of the crystal. 1 and 2 are indices relative to signal and idler fields, respectively.

The two-photon detection probability is proportional to the fourth-order correlation function \cite{mandel95}
\begin{equation}
P(\rrho_1,\rrho_2) \propto \bra{\psi}\oper{E}_1^\dagger(\rrho_1)\oper{E}^\dagger_2(\rrho_2)\oper{E}_1(\rrho_1)\oper{E}_2(\rrho_2)\ket{\psi}, 
\label{eq3}
\end{equation}
where $\oper{E}(\rrho)$ is the detection operator for a photon detected at position $\rrho$.  
Assuming that the pump beam is sufficiently weak so that the production of multiple photon pairs is negligible,  one can associate a wave function $\Psi$ to the two photon state so that $P(\rrho_1,\rrho_2) = |\Psi(\rrho_1,\rrho_2)|^2$, where \cite{Rubin94}
\begin{equation}
\Psi(\rrho_1,\rrho_2) = \bra{0}\oper{E}_1(\rrho_1)\oper{E}_2(\rrho_2)\ket{\psi}.  
\label{eq4}
\end{equation}

\begin{figure}
\centering
\includegraphics[width=0.75\columnwidth]{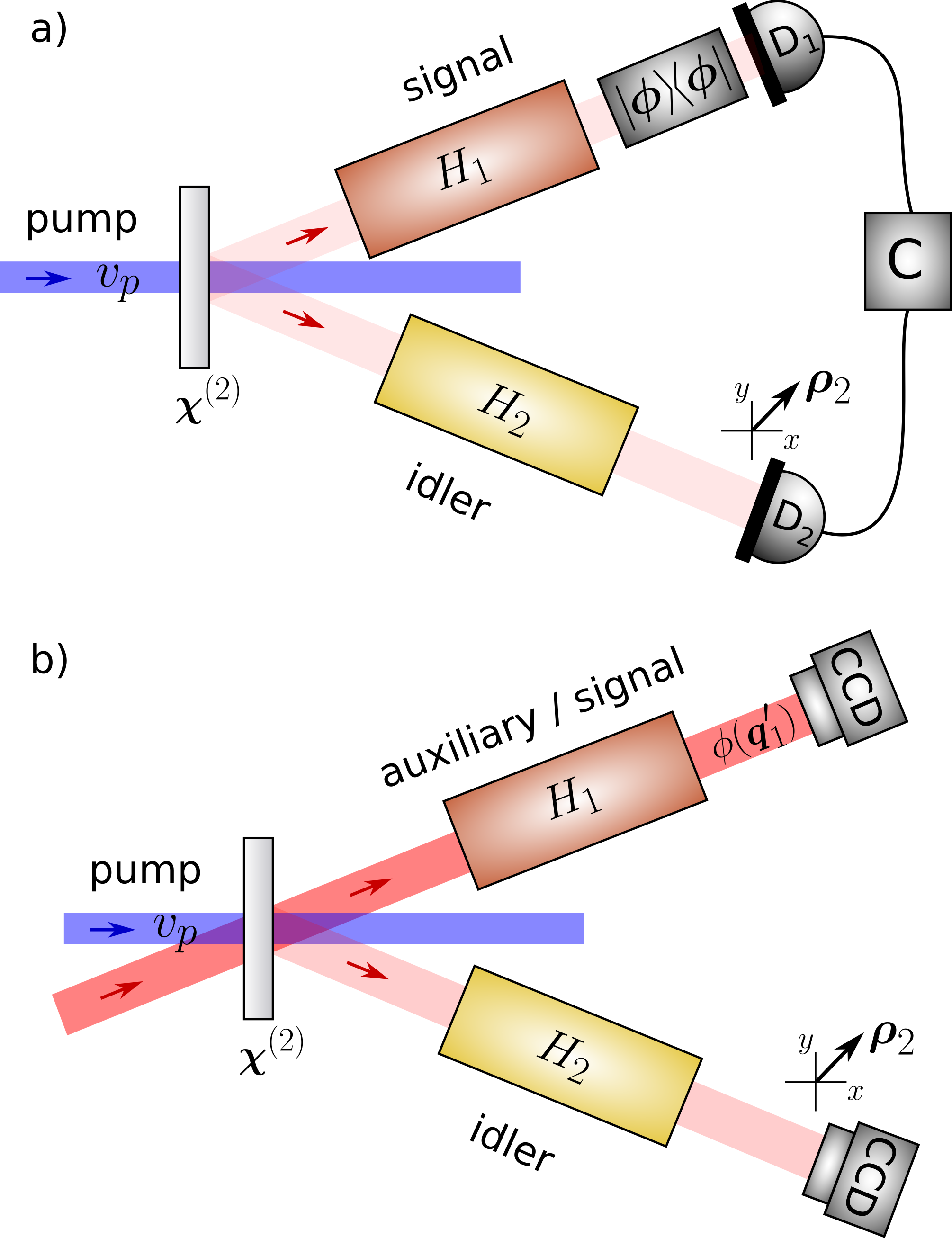}
\caption{a) Spontaneous PDC using a pump laser with spatial structure.  The down-converted photons travel through optical systems $H_1$ and $H_2$.  Photon 1 is projected onto spatial mode $\phi$. b) In the similar scheme for Stimulated PDC, an auxiliary laser in sent along the signal direction, stimulating generation of signal photons in the laser mode and conjugated twin photons in the idler mode.}
\label{fig:awp_sketch} 
\end{figure}

Two-photon coincidence imaging has been considered by a number of authors \cite{monken98a,abouraddy2001,abouraddy2002}. Let us suppose that photons 1 and 2 propagate through optical systems described by the transfer functions $H_1$ and $H_2$, as illustrated in Fig. \ref{fig:awp_sketch}a. We also assume that photon 2 will be detected by a point detector. In this case the detection operator is
\begin{align}
\oper{E}_2(\rrho_2)= \frac{1}{2\pi}\int d\qq_2 d\qq'_2\, e^{i\qq'_2\cdot \rrho_2}H_2(\qq_2,\qq'_2)\, \oper{a}_2({\qq_{2}}),
\label{eq:E2}
\end{align}
where the annihilation operator $ \oper{a}_2(\qq_2)$ annihilates a photon in the optical mode with transverse momentum component $\qq_2$.  

Consider now that photon 1 is projected onto the spatial mode $\phi$. There are a number of possible strategies allowing this kind of projection. One nice example is the projection onto a Laguerre-Gaussian mode using a single mode optical fiber and a holographic mask \cite{mair01}. In all cases, projection onto a spatial mode $\phi$ can be performed using an optical mode selector and a single mode fiber. The detection operator in this case is given by $\oper{E}_1(\rrho_1) \rightarrow \oper{E}_{1\phi}$, where
\begin{align}
\oper{E}_{1\phi}= \int d\qq_1 d\qq'_1 H_1(\qq_1,\qq'_1)\phi^*(\qq'_1)\oper{a}_1(\qq_1),  
\label{eq:E1}
\end{align}
and $\phi(\qq)$ is the mode's angular spectrum. 
Using operators \eqref{eq:E2} and \eqref{eq:E1} in 
\begin{equation}
\Psi_\phi(\rrho_2) = \bra{0}\oper{E}_{1\phi}\oper{E}_2(\rrho_2)\ket{\psi},
\label{eq5}
\end{equation} 
the two-photon wave function for a thin crystal becomes
\begin{align}
\Psi_\phi(\rrho_2) = &\frac{1}{2\pi} \int d\qq_1 d\qq_2 d\qq'_1
d\qq'_2 \, v(\qq_1+\qq_2) \nonumber  \\
& \, H_1(\qq_1,\qq'_1)\, H_2(\qq_2,\qq'_2)\, e^{i\qq'_2\cdot\rrho_2} \, \phi^*(\qq'_1). 
\label{eq:spdc}
\end{align}
The wavefunction \eqref{eq:spdc} closely resembles the output field \eqref{eq:in-out-E}. That is to say, the parametric down-conversion within the present approximations is isomorphic to an input-output classical optical system in the spirit of the AWP.

We can identify the angular spectrum $v(\qq_1+\qq_2)$ with the transfer function $v(\qq'-\qq'')$ of an optical element having an added reflection in the $\qq_2$ coordinate. Thus, the two photon state from SPDC in the case where one photon is projected onto the arbitrary state $\ket{\phi}$ is equivalent to the optical system shown in Fig. \ref{fig:spdc_awp}. The spatial structure of the pump beam comes into play through its angular spectrum $v$. In this analogy the transmission object described by $t$ is replaced with a reflective object given by the pump beam acting as a mirror $t(\qq'-\qq'') \rightarrow v(\qq'-\qq'')$.

\section{Stimulated emission}
\label{sec:stim}

The emission of photons in the PDC process can be enhanced using a coherent light field to stimulate, say, the signal field. This is often done by aligning an auxiliary laser along the same direction as the signal field (see Fig. \ref{fig:awp_sketch}b). This procedure enhances the creation of signal photons in the mode of the coherent laser (spatial, polarization and frequency) by stimulated emission, concomitantly creating the counterpart idler photons.

The state produced by the stimulated process, within the monochromatic, paraxial and thin-crystal approximations, is \cite{Wang90}
\begin{align}
\ket{\psi}=&\ket{v_s(\qq)}\ket{0}+ \nonumber\\
&C\Int d\qq_1 d\qq_2\ v_p(\qq_1+\qq_2)
\oper{a}^\dagger(\qq_1)\ket{v_s(\qq)}\ket{1;\qq_2},
\label{eq:psi_stim}
\end{align}
where 1 and 2 are indices for signal and idler, respectively, $v_s(\qq)$ is the angular spectrum of the stimulating field at $z=0$ (at the crystal) and $\ket{v_s(\qq)}$ is the corresponding multimode coherent state in the continuus mode representation (p. 565 of Ref. \cite{mandel95}).

The idler intensity at a distance $z$ from the crystal and transverse coordinates $\rrho_2$ is given by the second-order correlation function
\begin{equation}
\mathcal{I}(\rrho_2)=\langle \oper{E}^{(-)}_2(\rrho_2) \oper{E}^{(+)}_2(\rrho_2)\rangle\ ,
\end{equation}
where $\oper{E}^{(+)}_2(\rrho_2)$ is the field propagated through the optical system of transfer function $H_2$ between the crystal and the detection planes:
\begin{equation}
\oper{E}^{(+)}_2(\rrho_2)=\frac{1}{2\pi}\Int d\qq'_2 d\qq_2 H_2(\qq'_2,\qq_2)\oper{a}(\qq_2)
e^{i\qq'_2\cdot\rrho_2}\ .
\end{equation}
As the electric field operator applied to vaccum yields 0, only the second term in Eq. \eqref{eq:psi_stim} contributes to the intensity. Thus,
\begin{align}
\oper{E}^{(+)}_2(\rrho_2)\ket{\psi}\propto\ &\Int d\qq' d\qq_1 d\qq_2 
H_2(\qq_2,\qq')e^{i\qq'\cdot\rrho_2} \nonumber\\
&\times v_p(\qq_1+\qq_2)\oper{a}^\dagger(\qq_1)\ket{v_s(\qq)}\ket{0}
\end{align}
and, therefore,
\begin{align}
\mathcal{I}(\rrho_2)\propto\ &\Int d\qq'' d\qq'_1 d\qq'_2 d\qq' d\qq_1 d\qq_2 H_2^\ast(\qq'_2,\qq'')H_2(\qq_2,\qq') \nonumber\\
&\times e^{-i(\qq''-\qq')\cdot\rrho_2}v_p^\ast(\qq'_1+\qq'_2)v_p(\qq_1+\qq_2) \nonumber\\
&\times \bra{v_s(\qq)}\oper{a}(\qq'_1)\oper{a}^\dagger(\qq_1)\ket{v_s(\qq)}
\label{eq:I_integral}
\end{align}
The commutation relation for the bosonic operators yields $\bra{v_s(\qq)}\oper{a}(\qq'_1)\oper{a}^\dagger(\qq_1)\ket{v_s(\qq)}=
\delta(\qq'_1-\qq_1)+v_s^\ast(\qq_1)v_s(\qq'_1)$, splitting the integral of Eq. \eqref{eq:I_integral} in two parts:
\begin{equation}
\mathcal{I}(\rrho_2)=\mathcal{I}_\textup{spont}(\rrho_2)+\mathcal{I}_\textup{stim}(\rrho_2)\ .
\label{eq:two-parts}
\end{equation}
The first part -- the one involving $\delta(\qq'_1-\qq_1)$ -- describes the spontaneous emission, while the second one -- which carries $v_s^\ast(\qq_1)v_s(\qq'_1)$ -- represents the stimulated process. This nice and simple result was obtained in Ref. \cite{PRA-ImageCoh} for free propagation after the crystal and is generalized here for any given transfer function $H_2$. In order to prove this statement, we use the fact that the angular spectrum $v$ is the Fourier transform of the amplitude profile $\W$. With that, a straight-forward calculation provides the expression for the first term on the right-hand side of Eq. (\ref{eq:two-parts}):
\begin{align}
\mathcal{I}_\textup{spont}&(\rrho_2)\propto\Int d\rrho\,\left|\W_p(\rrho)\right|^2 \nonumber\\
&\times\left|\Int d\qq'_2 d\qq_2 \,H_2(\qq_2,\qq'_2)e^{i(\qq'_2\cdot\rrho_2-\qq_2\cdot\rrho)}\right|^2\ ,
\label{eq:spont-only}
\end{align}
which, on the grounds that it does not depend on the profile of the auxiliary laser, gives the total intensity due to the spontaneous emission after propagation through the optical system of transfer function $H_2$. 

It is worth noting that, in the case of paraxial free propagation from crystal to detector (Eq. \ref{eq:freeprop}), the second line in Eq. (\ref{eq:spont-only}) reduces to 1. This implies that the detected intensity after free propagation contains no information on the spatial structure of the pump beam, since the squared modulus of the pump profile $\W_p$ is integrated over the transverse spatial coordinates.

The second term on the right-hand side of Eq. (\ref{eq:two-parts}) can be written as:
\begin{align}
\mathcal{I}_\textup{stim}(\rrho_2)\propto\left|\Int d\qq'_2 \right. & d\qq_1 d\qq_2 
\, H_2(\qq_2,\qq'_2)e^{i\qq'_2\cdot\rrho_2}  \nonumber\\
& \times v_p(\qq_1+\qq_2)v_s^\ast(\qq_1)\bigg|^2 ,
\label{eq:stim-only}
\end{align} 
which gives the intensity due to StimPDC. We can see that it depends on $v_s^\ast(\rrho)$, the angular spectrum of the auxiliary beam in the crystal plane. The contribution from the stimulated emission can be made much stronger than that from spontaneous emission if the auxiliary laser intensity is high enough. Typically, a few milliwatts are enough to produce a stimulated emission 100 times stronger than the spontaneous emission.

To see the relation between StimPDC and the AWP, we note that the last equation is isomorphic to Eq. (\ref{eq:in-out-E}). Indeed, if we assume that the angular spectrum of the auxiliary field is prepared by sending an initial field $\phi^*(\qq'_1)$ back through an optical system represented by the transfer function $H_1$, we can replace
\begin{equation} 
v_s^\ast(\qq_1)=\int d\qq'_1 H_1^B(\qq'_1,\qq_1) \phi^\ast(\qq'_1)
\label{eq:replace}
\end{equation}
in Eq. (\ref{eq:stim-only}), where $H_1^B$ is the transfer function of the optical system $H_1$ in the backwards direction (from the detector to the crystal). Then, we get
\begin{align}
\mathcal{I}_\textup{stim}(\rrho_2)\propto & \left|\Int d\qq'_1 d\qq'_2 d\qq_1 d\qq_2 \right. \,
e^{i\qq'_2\cdot\rrho_2}\phi^\ast(\qq'_1)\nonumber\\
& \times  H_1^B(\qq'_1,\qq_1) v_p(\qq_1+\qq_2) H_2(\qq_2,\qq'_2) \biggr|^2,
\label{stim:awp}
\end{align} 
which is identical in form to Eq. (\ref{eq:in-out-E}). We conclude that the AWP, in this sense, also applies to StimPDC.

The auxiliary laser field $v_s^{\ast}(\qq_1)$ can be properly prepared using a spatial light modulator (SLM) for instance, in order to help in the design of experiments with twin photons from SPDC. Moreover, it can be helpful for alignment, since one can align the set-up with the auxiliary laser and the stimulated idler beam, which is so intense that we can observe it with a common and inexpensive CCD camera, or even the naked eye, depending on the wavelength. Once the set-up is aligned, one can just turn off the auxiliary laser and perform coincidence counting experiments.

In the next section, we present a number of examples of how this can be used for novel optical experiments.    

\section{Examples and Experiments with stimulated emission}
\label{sec:experiments}

\begin{figure}
\centering
\includegraphics[width=\columnwidth]{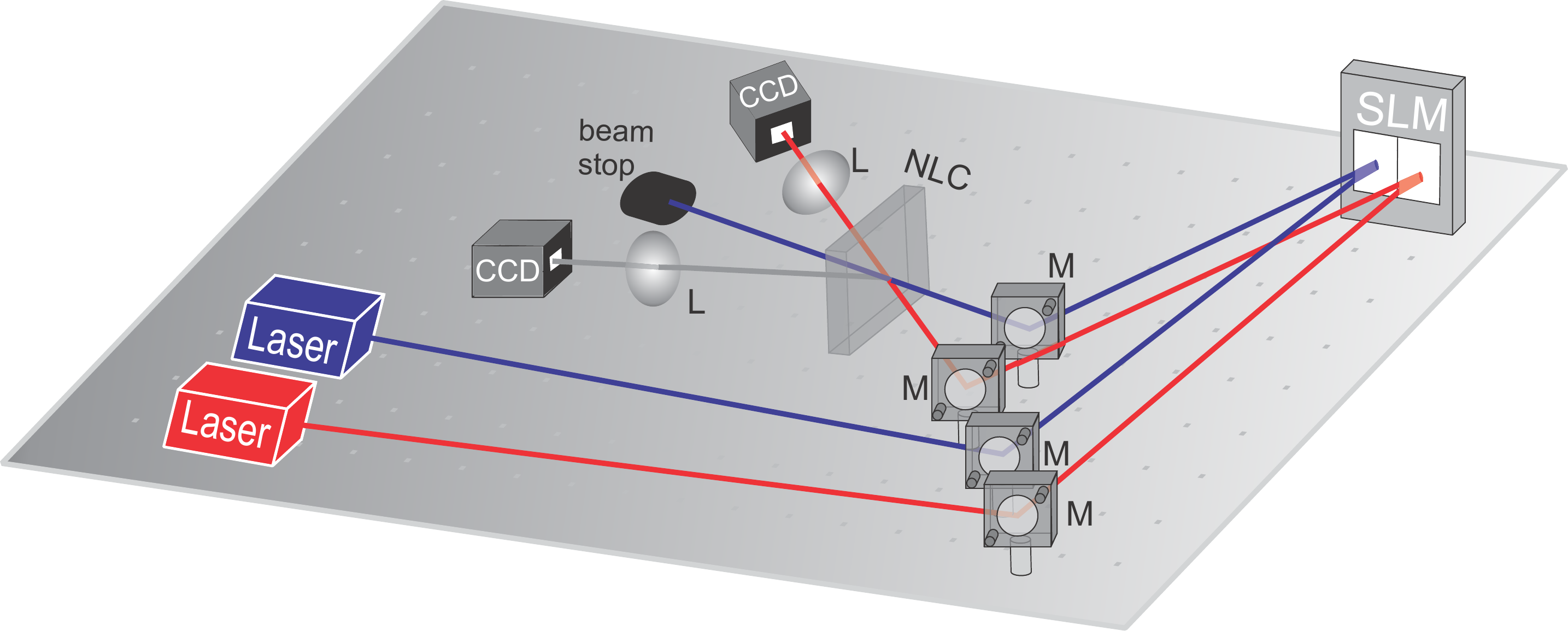}
\caption{General set-up used for our StimPDC experiments.}
\label{fig:setup} 
\end{figure}

We illustrate the usefulness of stimulated parametric down-conversion in the design of experiments with spontaneous emission. The experimental set-up is sketched in Fig. \ref{fig:setup}. A diode laser oscillating at 405 nm pumps a BBO nonlinear crystal. We work in a non-collinear phase-matching configuration with very small angle between signal and idler, $~1^\circ$, and use 10 nm bandwidth interference filters centered at 780 nm (signal) and 840 nm (idler). Another diode laser (auxiliary laser) at 780 nm is aligned with the signal direction and stimulates the emission in both signal and idler. Pump and auxiliary lasers can be spatially modulated on demand, which is represented on the picture by reflection on an SLM. 

The intensity of the emission in the idler field is strongly enhanced in comparison to the case of only spontaneous emission, generating a beam with macroscopic intensity whose transverse profile is detected by a CCD camera, along with a relatively small background coming from spontaneous emission. In practice, we can monitor the stimulated emission profile in real time. The auxiliary laser transverse profile is also monitored with a CCD camera.

\subsection{Phase conjugation effects}

The physical phenomenon behind the AWP is phase conjugation. For time-dependent waves, phase conjugation is equivalent to temporal reversion. Therefore, the advanced wave (as in the AWP) is exactly the time reversal of the signal beam. In stimulated down-conversion,
the phase conjugation is evident in Eq. (\ref{eq:stim-only}). It is the conjugate of the auxiliary beam's angular spectrum that, together with the pump, determines the properties of the idler beam. For instance, if the pump beam has a flat transverse field distribution, the idler beam will propagate forward as if it were the reflection of the advanced wave (signal propagating backwards). Indeed, with a flat pump profile, the angular spectrum is peaked around 0, $v_p(\qq_1+\qq_2)=\delta(\qq_1+\qq_2)$; plus, considering a transfer function $H_z$ representing free propagation over a distance $z$, Eq. (\ref{eq:stim-only}) gives:

\begin{equation}
\mathcal{I}_\textup{stim}(\rrho_2)\propto\left|\Int d\qq'_2 d\qq_2 
\, H_z(\qq_2,\qq'_2)v_s^\ast(-\qq_2) e^{i\qq'_2\cdot\rrho_2}\right|^2,
\end{equation}
where the negative sign in $v_s^\ast(-\qq_2)$ indicates the mirror reflection and the integration over $\qq'_2$ performs the Fourier transform from $\qq$-space to $\rrho$-space.

\begin{figure}
 \centering
 \includegraphics[width=0.8\columnwidth]{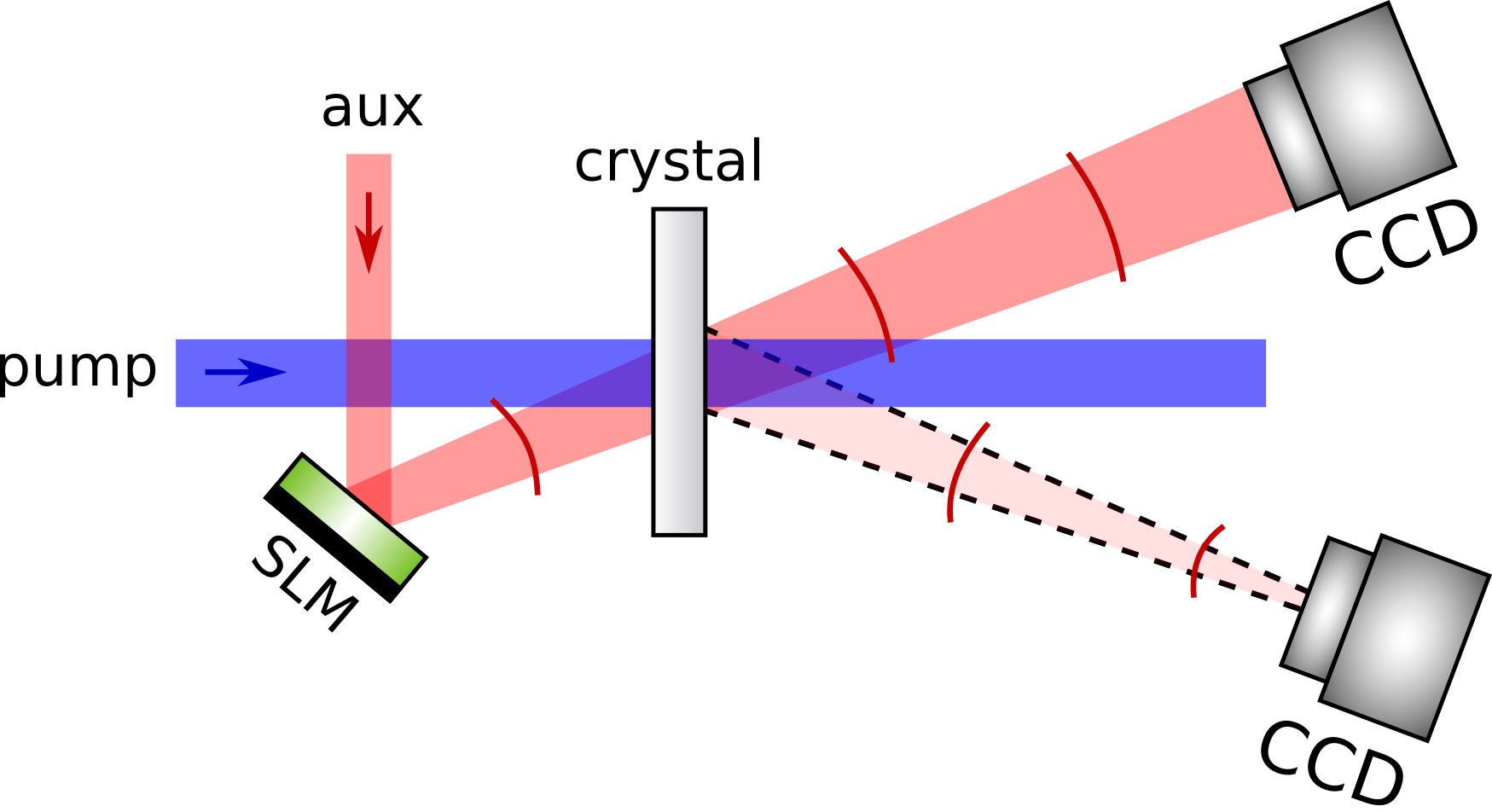}
 \caption{Experimental scheme for phase conjugation (flat pump wave front).}
 \label{fig:flat} 
 \end{figure}

\begin{figure}
\centering
\includegraphics[width=0.8\columnwidth]{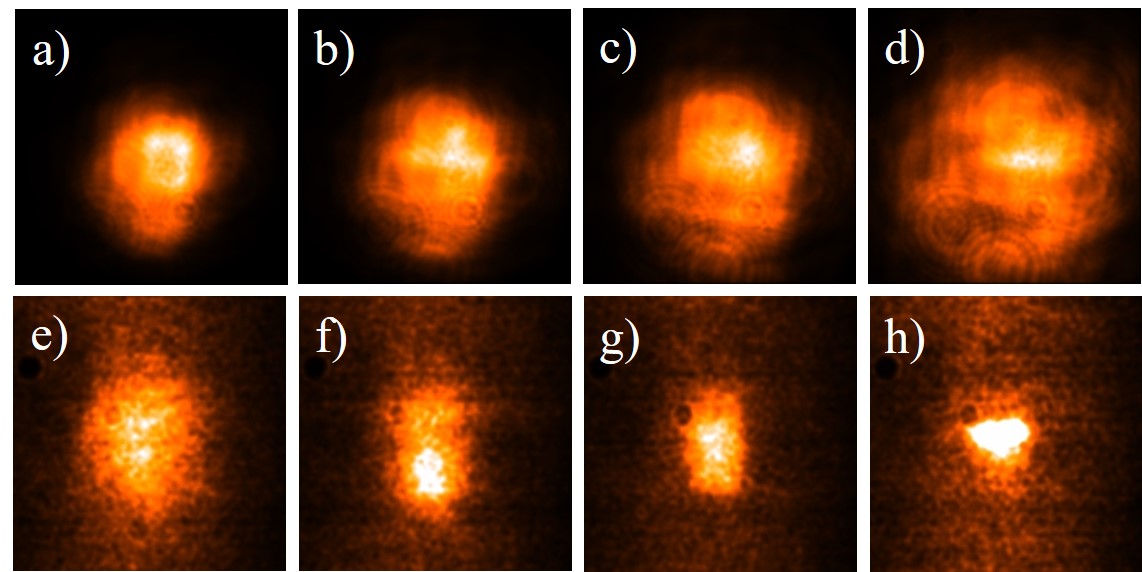}
\caption{Experimental results for phase conjugation. a)-d) Increasing spot size of the auxiliary beam registered by the CCD camera as we make it diverge. e)-h) Corresponding decreasing idler spot size. At the background, one can see a portion of the spontaneous emission cone.}
\label{fig:pconj_results} 
\end{figure}

\textbf{Experiment.} In reference \cite{PRL-PhaseConj}, the authors observed effects of phase conjugation by analyzing the symmetry of images transferred from the pump and auxiliary beams to the idler. Here, we illustrate the phase conjugation effects by observing the focusing of the idler beam as we make the auxiliary beam diverge. The scheme is sketched in Fig \ref{fig:flat}. The pump is collimated, so that its wave front and amplitude distribution are practically flat. The auxiliary laser is sent to the SLM, where a divergent lens of variable focal length is implemented, and its profile is monitored with a CCD camera. We start with very long focal length, so that the auxiliary beam stays collimated. In this case, the idler beam has its largest spot size. As we force the auxiliary laser to diverge (by bringing the focal length from $-\infty$ toward 0), its spot size in the camera increases. Consequently, the idler spot starts decreasing. That is to say, because the idler beam reproduces the auxiliary's advanced wave reflected by a flat mirror, its spot size varies in the opposite way as compared to the auxiliary. If the auxiliary field is converging, the idler is diverging, and vice-versa. We have observed this behavior, and the results are shown in Fig. \ref{fig:pconj_results}.


\subsection{Phase-modulated pump - Fractional Fourier Transform}

The Fractional Fourier Transform (FRFT) has been studied in the context of coincidence imaging by several authors \cite{tasca08,tasca09a,tasca09b,Liu09,Wang09,kang10,sun14}.  In these experiments, the down-converted photons pass through optical systems that are used to implement the FRFT.  Here we show that the FRFT can be implemented in the AWP by controlling the pump beam alone.  The FRFT is a generalization of the usual Fourier transform, and it is parameterized by the angle $\alpha$.  Its kernel is given by
\begin{equation}
F_{\alpha}(\qq,\qq^\prime)\propto \exp\left \{ \frac{i}{2}\left [ \cot \alpha(\qq^2 +\qq^{\prime \, 2}) - 2\frac{\qq^\prime\cdot\qq}{\sin \alpha} \right] \right\},
\label{eq:frft1}
\end{equation}  
which transforms a function $\Psi(\qq)$ into a function $\Phi(\qq')$. Here $\qq$ and $\qq'$ are dimensionless variables, which can be represented by different coordinate axes in a bosonic phase space.
For $\alpha=\pi/2$, one recovers the usual Fourier transform:
\begin{equation}
F_{\frac{\pi}{2}}(\qq,\qq')\propto \exp\left ( - i \,\qq'\cdot\qq \right ),
\label{eq:fr}
\end{equation}  
where $\qq$ and $\qq'$ are canonically conjugate. For $\alpha=\pi$, the FRFT kernel corresponds to:  
\begin{equation}
F_{\pi}(\qq,\qq^\prime)\propto \delta(\qq+\qq^\prime),   
\label{eq:Im}
\end{equation}  
where $\qq$ and $\qq^\prime$ live in the same space, which can be understood in the context of optics as an imaging system.

Let us now return to the wavefunction \eqref{eq:spdc} and consider that the pump beam is a Gaussian beam with angular spectrum given, up to a constant phase factor, by: 
\begin{equation}
v_p(\qq) = f_p(\qq)\,\exp\left(-i\frac{Z}{2K}\qq^2\right),
\label{eq:gaussbeam}
\end{equation}
where $K=2\pi/\lambda_p$ is the pump beam wavenumber, $f_p$ is the Gaussian envelope defined at the waist position, and $Z=Z_C-Z_0$ with $Z_C$ and $Z_0$ being, respectively, the positions of the non-linear crystal and the pump beam waist along the propagation axis. For the implementation of the FRFT, we consider the special case where signal and idler photons are frequency degenerate with wavenumber $k=2\pi/\lambda=K/2$, and propagate freely over the same distance $z$ between the crystal and the corresponding detection planes. Using the free propagation transfer functions for signal and idler given in Eq. (\ref{eq:freeprop}), we write the two-photon wavefunction at the detection planes as 
\begin{align}\label{eq:Propwavefunction}
\Phi(\qq_1,\qq_2) & =  f_p(\qq_1+\qq_2)  \\ \nonumber
& \times \exp{\left\{-i\left[\frac{Z(\qq_1+\qq_2)^2}{2K}+\frac{z(\qq_1^2+\qq^2_2)}{2k} \right]\right\}} . \\ \nonumber
\end{align}
The propagated two-photon wavefunction \eqref{eq:Propwavefunction} bears close similarity with the FRFT kernel given in Eq. \eqref{eq:frft1}. The proper identification between them require the definition of dimensionless variables $\bq_1 = s \qq_1$ and $\bq_2 = s \qq_2$, with `$s$' being a constant with dimension of length. We further identify the angle $\alpha$ such that 
\begin{align}\label{FRFT_def}
\frac{Z}{K s^2} = \frac{1}{\sin \alpha},  \quad \textrm{and} \quad \left(\frac{z}{k}+\frac{Z}{K}\right)\frac{1}{s^2} = -\cot \alpha.
\end{align}
Using the definitions \eqref{FRFT_def}, the propagated two-photon wavefunction \eqref{eq:Propwavefunction} in terms of the dimensionless variables $\bq_1$ and $\bq_2$ can be written as a function of the quadratic phase term associated with the FRFT kernel given in Eq. \eqref{eq:frft1} :
\begin{equation}
\Phi(\bq_1,\bq_2)  =  f_p\left(\frac{\bq_1+\bq_2}{s} \right) F_{\alpha}(\bq_1,\bq_2) .
\end{equation}

Similarly to what discussed in section \ref{sec:QS}, we consider the mode of photon $2$, $\Phi_\phi(\qq_2)$, produced by the projection of photon $1$ onto the spatial mode $\phi(\qq_1)$:
\begin{equation}\label{Phi2}
\Phi_\phi(\bq_2)  =  \int d\bq_1 \Phi(\bq_1,\bq_2)  \phi^*\left( \bq_1/s \right).
\end{equation}
We will also assume that the pump beam envelope function $f_p$ is much broader than the mode function $\phi$, so that $f_p$ becomes essentially a constant factor in the integrand given in Eq. \eqref{Phi2}. We finally write the mode function of photon $2$ as
\begin{equation}\label{FRFTphi2}
\Phi_\phi(\bq_2) \propto   \int d\bq_1 \phi^*\left(\bq_1/s \right) F_\alpha(\bq_1,\bq_2)  \propto  \mathcal{F}_\alpha \{ \phi^*\left(\bq_1/s \right)\} ,
\end{equation}
where $\mathcal{F}_\alpha \{ \cdot \}$ denotes the FRFT of order  $\alpha$. We then conclude that the mode function describing photon $2$ is given by the FRFT of the mode $\phi^*(\qq)$ in which the photon $1$ is projected. Similarly, the detection amplitude for photon $2$ [such as in Eq. \eqref{eq:spdc}] can be written as a FRFT in position representation by making use of $\psi^*(\rrho_1)$, the Fourier transform  $\phi^*(\qq)$:
\begin{equation}
\Psi_\phi(\bar{\rrho}_2)\propto \int d\bq_2  \Phi_\phi(\bq_2) e^{i \bq_2\cdot \bar{\rrho}_2} \propto \mathcal{F}_\alpha \{ \psi^*(s \bar{\rrho}_1 )\},
\label{eq:frft4}
\end{equation}
where we also define dimensionless variables in position representation: $\bar{\rrho}_i=\rrho_i/s$ ($i=1,2$). 

In order to gain intuition on the order $\alpha$ of the FRFT as a function of the experimental parameters, we solve Eqs. \eqref{FRFT_def} to get
\begin{equation}
\cos \alpha=-\left( 2\frac{z}{Z}+1\right),
\label{eq:cosalpha}
\end{equation}
which provides $\alpha$ as a function of the longitudinal position of the detection planes ($z$) and the pump beam waist ($Z$) in relation to the non-linear crystal. Note that Eq. \eqref{eq:cosalpha} imposes a constraint on the ratio $z/Z$ for the implementation of lensless FRFT. First of all, we note that $Z=Z_C-Z_0<0$, \textit{i.e.}, the pump beam must be convergent at the non-linear crystal, so that its waist is located \textit{after} it. This is in fully agreement with an advanced wave picture of this FRFT implementation: the phase curvature of the pump must act as a \textit{concave} mirror for the advanced wave $\phi^*(\qq)$ `emitted' from detector's $1$ location. A careful inspection of Eq. \eqref{eq:cosalpha} provides the prescription on the location of the pump beam waist within the range $Z_0\geqslant Z_C+z$ for a particular choice of $\alpha$. For instance, working with $Z=-z$ gives $\alpha=0$ while $Z=-2z$ gives $\alpha=\pi/2$ (the standard Fourier transform). In the more general case, proper control of the phase curvature of the pump beam allows for lensless implementation of the FRFT, which has applications in imaging and signal processing \cite{ozaktas01}.  We note that a lensless implementation of the FRFT was performed using partially coherent light and coincidence detection in Ref. \cite{Wang09}, though phase curvature of the beam was not relevant in that case.

A number of previous results and possible applications can be obtained as special cases of Eqs. \eqref{eq:frft4} and \eqref{eq:cosalpha}.  For example, the imaging experiment performed by Pittman et al. \cite{pittman96a} is obtained from Eq. \eqref{eq:frft4} by choosing $\alpha=\pi$. 
Recalling that $\phi^*(\qq)$ represents a certain transverse mode preparation of the AWP light source, its Fourier transform is actually the image of some aperture function
(which implements the mode filtering) placed in front of the detector. We note that the a similar effect was used to optimize the pair collection efficiency in SPDC \cite{monken98b}.

The AWP-equivalence between SPDC (Eq. \ref{eq:spdc}) and StimPDC (Eq. \ref{eq:stim-only}) suggests that we have, for StimPDC:
\begin{align}
\mathcal{I}_\textup{stim}(\rrho_2) 
\propto \left|\Psi_\phi(s{\rrho}_2)\right|^2,
\label{eq:int-frft}
\end{align}
where $\Psi_\phi$ is given by Eq. (\ref{eq:frft4}) and $\phi^*$ is prepared with the stimulating laser according to Eq. (\ref{eq:replace}).

\begin{figure}
 \centering
 \includegraphics[width=0.8\columnwidth]{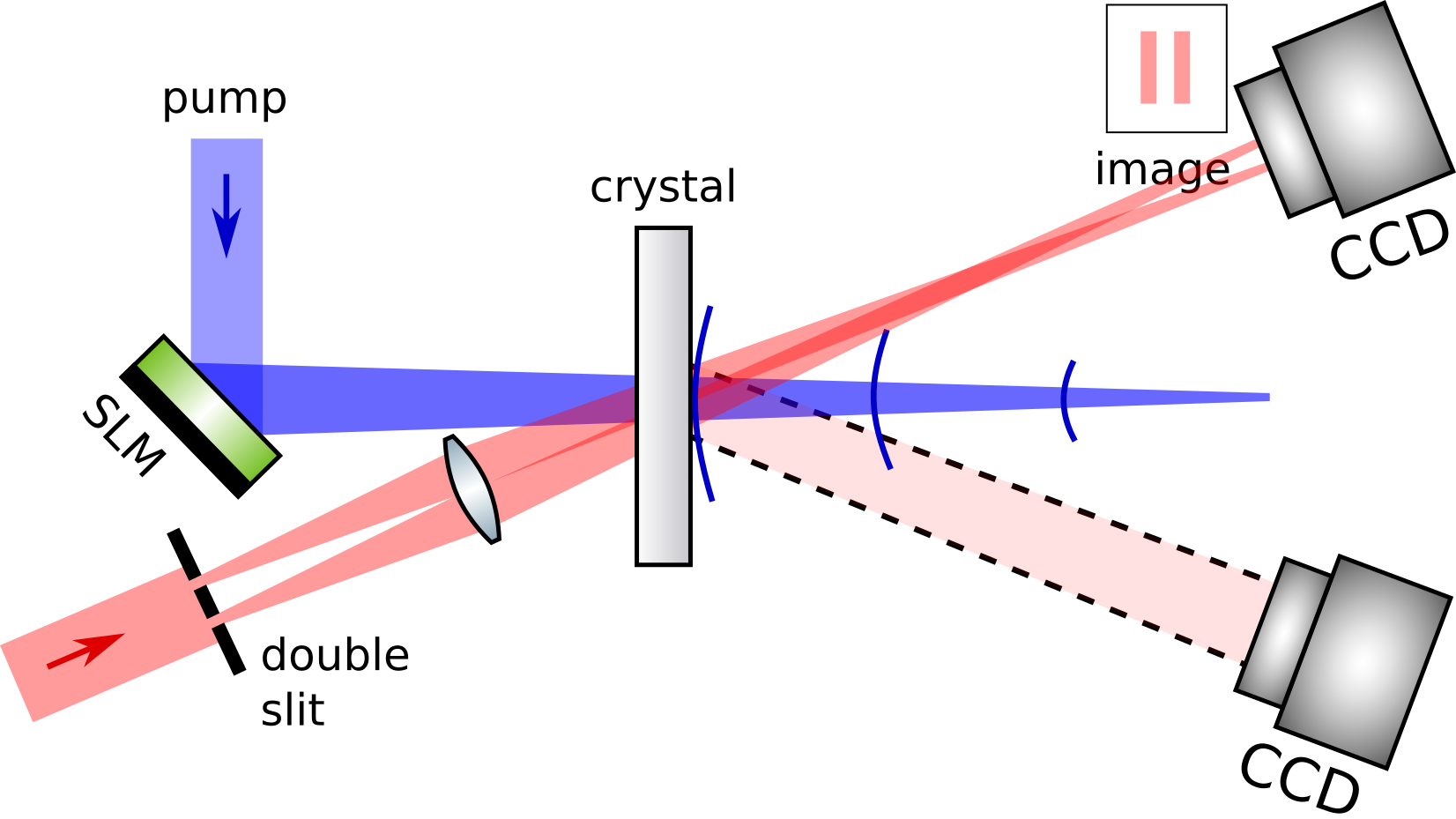}
 \caption{Experimental scheme for Fractional Fourier Transform. The curved pump wave front acts as a lens, which performs a FRFT on the auxilary beam's advanced wave.}
 \label{fig:phase}
 \end{figure}

\textbf{Experiment.} We now illustrate the effect of phase modulation of the pump transverse profile in stimulated emission by performing the optical FRFT of the auxiliary field, observed in the idler profile. Consider that the angular spectrum $\phi(\qq)$ of the field
\begin{equation}
\W(\rrho)\propto\left\{
\begin{array}{ll}
1& \quad\textup{if }\tfrac{d-\delta}{2} < |x| < \tfrac{d+\delta}{2}\\ 
0& \quad\textup{elsewhere}\ 
\end{array},
\right.
\label{eq:doubleslit}
\end{equation}
which characterizes the image of a double slit, centered at $x=0$, with slit width $\delta$ and separation $d$. For this kind of AWP source and performing the FRFT, from Eqs. (\ref{eq:int-frft}), (\ref{eq:doubleslit}) and (\ref{eq:angular-spec}) we expect to see transverse distributions varying between the usual Young double slit interference pattern when the FRFT parameter $\alpha = \pi/2$ and the image of the double slit when $\alpha = 0$. The FRFT parameter $\alpha$ is varied by changing the curvature of the pump beam wave front. We have done this with a variable focal-length lens implemented by an SLM (Fig. \ref{fig:phase}).


\begin{figure}
\centering
\includegraphics[width=\columnwidth]{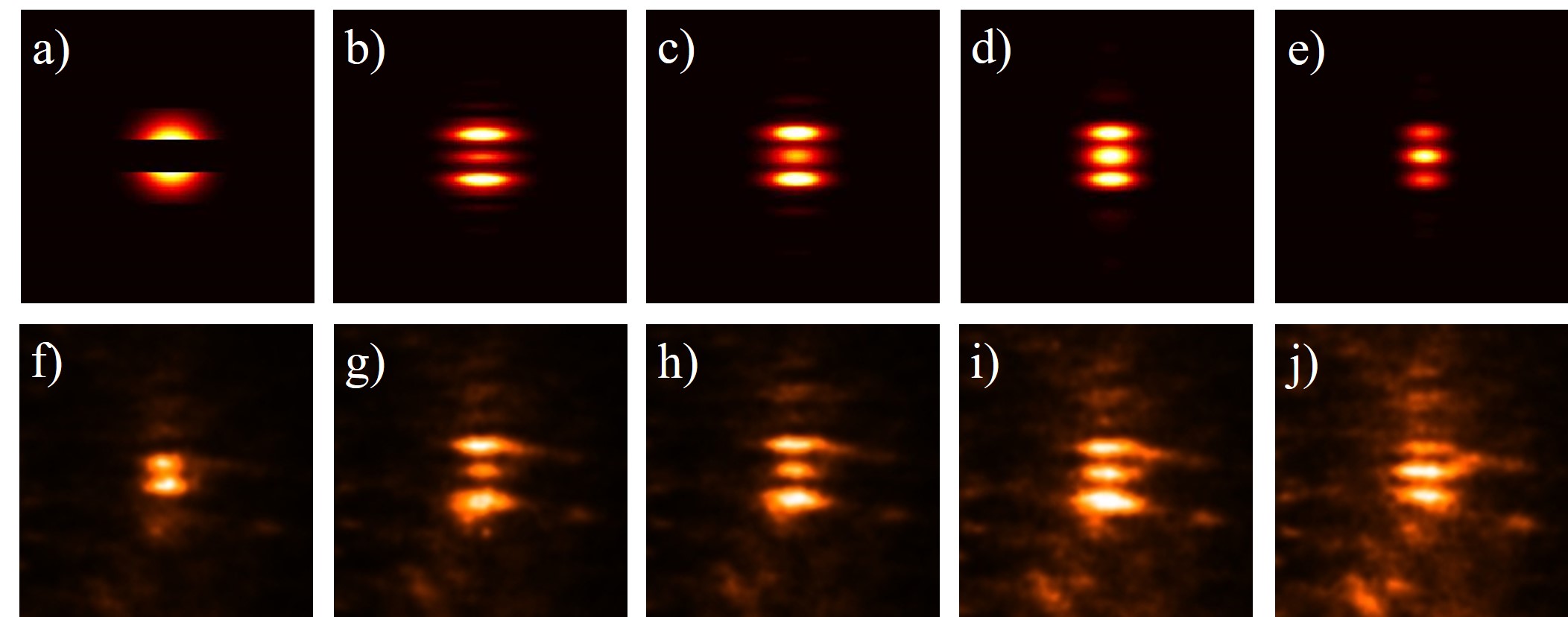}
\caption{Experimental results for the Fractional Fourier Transform. a)-e) Calculated FRFT of a Gaussian profile partially blocked by a double slit for several values of $\alpha$ (0, 0.20$\pi$, 0.25$\pi$, 0.30$\pi$ and 0.35$\pi$, respectively), using Eqs. (\ref{eq:int-frft}), (\ref{eq:doubleslit}) and (\ref{eq:angular-spec}). f)-j) Idler intensity profiles showing the corresponding experimental implementations of FRFT.}
\label{fig:frft_results} 
\end{figure}

The function $\phi$ is realized by sending the auxiliary laser through a double slit, which is imaged to the detection plane by a lens located between the double slit and the crystal. The results are shown in Fig. \ref{fig:frft_results}, which compares, for different values of $\alpha$ implemented, the observed intensity of the idler beam with the simulated FRFT of the imaged auxiliary laser. Thus, our results show that the curvature of the pump field can be used to implement the optical FRFT of the transverse spatial amplitude of the auxiliary laser. Since the FRFT has found use in filtering and signal processing, our results could be interesting for these fields.


\subsection{Amplitude-modulated pump}

It is pedagogical to analyze the cases where the structure of the pump beam is transferred to the coincidence counting rate in the SPDC case, or to the idler intensity in the StimPDC case. Let us consider the situation where the pump beam propagates through some diffraction aperture, which is imaged onto a plane situated at distance $z$ after the crystal. Let us assume that the distances from the detection planes and the crystal are also $z$. It was shown in Ref. \cite{monken98a} that the coincidence counting rate is given by:
\begin{equation}
C(\rrho_1=0, \rrho_2) \propto |\W_p(\rrho_2/2, z)|^2,
\label{eq:transf}
\end{equation}
where $\W_p$ is the pump beam field profile at a distance $z$ from the crystal and $\rrho_1$ is fixed to $0$. From the AWP perspective, we have the signal detector acting as point source located at a distance $z$ from the crystal and $\rrho_1 = 0$. In this ideal case, we ignore the effects due to the finite size of the detector aperture. It emits an advanced wave which, after reflection in a structured mirror, acquires the angular spectrum of the pump and propagates over a distance $z$ to form the image given by $\W_p$. The only difference is that, due to the distinct wavelength, the coincidence image is two times larger than the actual pump profilem, as can be seen from the factor 1/2 in the argument of $\W_p$.

The equivalent scheme for stimulated down-conversion is obtained by replacing the signal point detector with the auxiliary laser focused on a plane at a distance $z$ from the crystal. In the plane of the crystal, the auxiliary laser will have the same angular spectrum as the advanced wave coming from a point source, neglecting the effects due to the finite size of the laser in the focal plane. Approximating the angular spectrum of the auxiliary laser in the crystal by a plane wave, with $v_s(\qq) = \delta(\qq)$, Eq. (\ref{eq:stim-only})
becomes:
\begin{align}
\mathcal{I}_\textup{stim}(\rrho_2) \propto  &\ \biggr| \int d\qq_2 \,\,  v_p(\qq_2)  
 \exp\left[i\left(\qq_2\cdot\rrho_2 -\frac{q_2^2}{2k_2}z\right)\right] \biggr|^2 \nonumber  \\ 
 = &\ \left|  \W_p(\rrho_2,z)  \right|^2.
\label{eq:struct}
\end{align}
Comparing Eqs. (\ref{eq:transf}) and (\ref{eq:struct}), we can see that in both cases the angular spectrum of the pump beam is transferred and, after propagation, the same image as the pump is measured in the idler side. The only difference is the scaling factor of 2 that appears only in the argument of $\W_p$ in Eq. (\ref{eq:transf}).

\textbf{Experiment.} We present here a slightly different case, where an obstacle (a thin horizontal and/or a vertical wire) is placed in front of the pump and imaged onto the crystal plane with a $4f$ imaging system. In this way, when the pump reaches the crystal, its amplitude shows the exact shape of the obstacle and we have a purely amplitude-modulated pump.

The auxiliary beam is collimated and sent to the crystal. The idler's intensity then results from the advanced wave propagating back from the detector to the crystal, ``reflecting'' in the amplitude-modulated pump and propagating towards the detection plane. In our specific experiment, the detection plane is about 30 cm behind the crystal, meaning that the advanced wave goes through some diffraction prior to detection.


 \begin{figure}
 \centering
 \includegraphics[width=0.8\columnwidth]{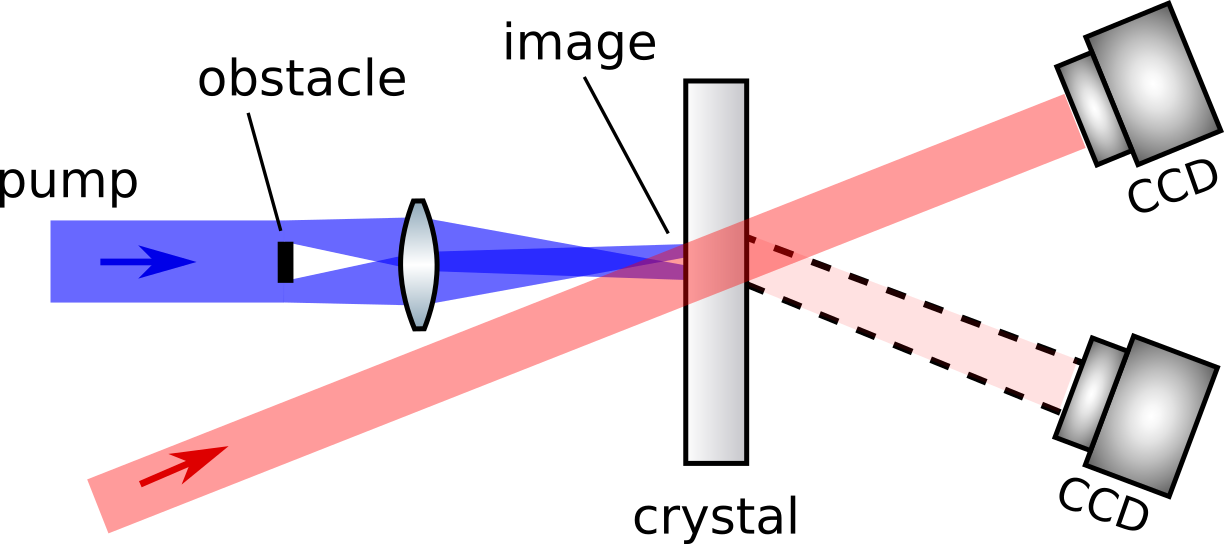}
 \caption{Experimental scheme for the amplitude-modulated pump, where the pump amplitude profile is transferred to the idler, given a flat auxiliary beam profile.}
 \label{fig:amplitude}
 \end{figure} 

\begin{figure}
\centering
\includegraphics[width=0.8\columnwidth]{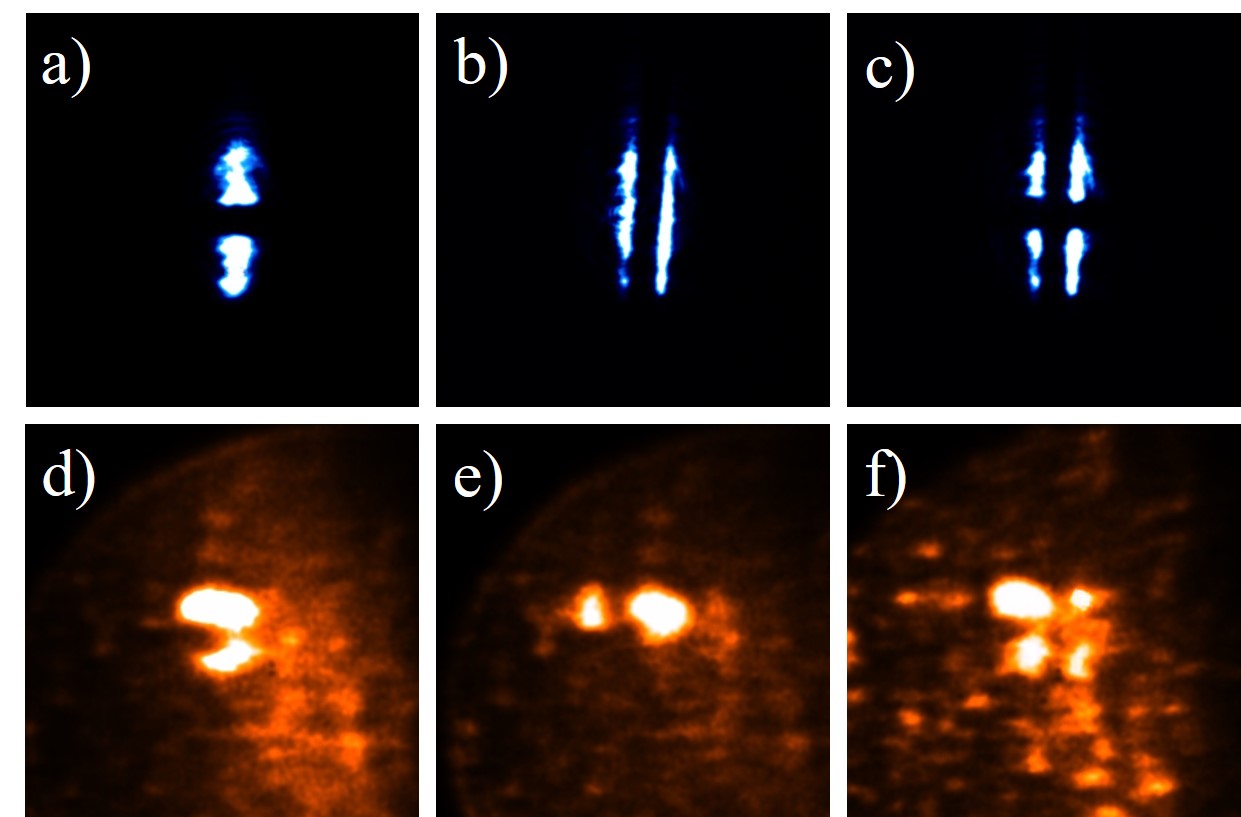}
\caption{Experimental results for amplitude-modulated pump. The pump intensity profile at the crystal plane is shown when we place a verticle wire (a), a horizontal wire (b) or two crossed wires (c). From d) to f), we see the corresponding idler intensity profiles which reproduce the pump amplitude modulation.}
\label{fig:wire} 
\end{figure}

Fig. \ref{fig:amplitude} illustrates the experimental setup and Fig. \ref{fig:wire} shows the result of the stimulated beam intensity, which displays approximately the same shape of the obstacle in front of the pump. The verification of Eq. (\ref{eq:struct}) is therefore done by observation of qualitative agreement between the measurements of $|\W_p(\rrho_2,z)|^2$ and $\mathcal{I}_\textup{stim}(\rrho_2)$.


\section{Using StimPDC to design SPDC experiments}
\label{sec:design}

An issue related to the design and realization of experiments involving pairs of photons and transverse spatial effects concerns the measurement of the coincidence patterns. The first approach at hand is to scan the photon detectors through the detection planes using small pinholes or optical fibers and reconstruct the conditional spatial structure. This task can be very time consuming, and becomes prohibitive if one needs to follow some iterative procedure to align or optimize parameters of the setup. The second possibility is to use intensified CCD cameras that allow the direct measurement of the whole two-photon coincidence \cite{Edgar2012,Aspden2013} pattern. However, even though this is more efficient than scanning, it also requires some time due to the weak flux of photons in the spontaneous parametric down-conversion in addition to other technical limitations of these devices. We propose and demonstrate here an alternative approach using the AWP applied to stimulated emission for testing and aligning such experimental set-ups.

Let us use the results discussed in section \ref{sec:experiments} to exemplify the use of StimPDC to design SPDC experiments. We have performed the experiment where an opaque object was used to modulate
the amplitude of the pump laser in the crystal. This was done in such a way that the image of the object was formed in a plane situated at a distance $z$ from the crystal. The auxiliary stimulating laser was just focused in plane, also at a distance $z$ from the crystal. As a
result, the intensity profile of the stimulated idler beam at a distance $z$ from the crystal had the same shape as the pump laser at a distance $z$ from the crystal.

Now, the SPDC version is easily obtained by turning off the auxiliary laser and detecting the signal beam in a way that its advanced wave reproduced the auxiliary laser propagation backwards the crystal. The auxiliary laser was just focused onto a plane. Therefore, the corresponding detection scheme for the signal beam is to use a small pinhole in front of the detector situated at a distance $z$ from the crystal. As a result, the coincidence transverse profile obtained scanning the idler detector also at a plane situated at a distance $z$ from the crystal will reproduce the pump beam intensity profile at the plane situated at a distance $z$ from the crystal. This is exactly what the measurements in Fig. \ref{fig:wire} illustrate for the StimPDC intensity.

\section{Conclusion}
\label{sec:conclusions}
   
Klyshko's advanced-wave picture is an extremely useful tool for understanding and designing two-photon coincidence experiments. Here we studied the advanced-wave picture considering a spatially structured pump beam in the context of both spontaneous and stimulated PDC. We show that, when the pump beam angular spectrum is properly prepared, it works in analogy to a spatial light modulator,
rather than as a simple mirror as described in the original version of the advanced-wave picture. This allows for a number of interesting applications in quantum imaging and the preparation of spatially entangled photons. Though the AWP has found widespread use in analyzing two-photon coincidence experiments (SPDC), we believe that this is the first time it has been applied to StimPDC. Linking the advanced-wave picture, typically used in analyzing correlations in SPDC, to StimPDC, which can be observed using a simple CCD camera or even the naked eye, suggests that StimPDC can be used to help design, build and align SPDC experiments. We discussed how the fractional Fourier transform can be performed in a quantum imaging scenario using the phase curvature of the pump laser as the lens, and presented an experimental implementation using StimPDC. We also show how the advance wave picture applied to both the spontaneous and stimulated cases describe the transfer of the angular spectrum from the pump to the down-converted fields.
We believe that this study can be helpful in the manipulation of the spatial correlations of twin photons from parametric down-conversion in several applications, as well in design and alignment of coincidence experiments.

\begin{acknowledgments}
The authors acknowledge financial support from the Brazilian funding agencies CNPq, CAPES, FAPERJ, FAPEAL and the National Institute for Science and Technology - Quantum Information.  
\end{acknowledgments}


\end{document}